\documentclass[aps,pre,preprint,groupedaddress]{revtex4}
\usepackage{graphicx}

\begin{document}

\title{System size stochastic resonance in driven finite arrays of coupled bistable elements}

\author{
Jos\'e G\'omez-Ord\'o\~nez, Jos\'e M. Casado, and Manuel Morillo}
\email{morillo@us.es}
\affiliation{Universidad de Sevilla. Facultad de F\'{\i}sica. \'Area de
F\'{\i}sica Te\'orica. Apartado de Correos 1065. Sevilla 41080. Spain}
ḉ

\date{\today}

\begin{abstract}
  The global response to weak time periodic
  forces of an array of noisy, coupled nonlinear systems might show a nonmonotonic dependence on the number of units in the array. This
  effect has been termed system size stochastic resonance. In this paper, we analyze the nonmonotonic dependence on the system size of the signal-to-noise ratio of a collective variable characterizing a finite array of one-dimensional globally coupled bistable elements.  By contrast with the conventional nonmonotonic dependence with the
  strength of the noise (stochastic resonance), system size stochastic resonance is found to be restricted to some regions in parameter space.
 \end{abstract}


\maketitle

\section{Introduction}
In some nonlinear systems, noise can be used to improve the response of the
system to a weak external signal. In the last two decades, this rather counterintuitive effect known
as Stochastic Resonance (SR), has been massively studied, both experimentally
and theoretically,  due to its potential interest to a
variety of areas of science \cite{GHJM1998} and technology \cite{BM2005}. A large majority of the experimental as well as theoretical studies have dealt mainly with the
response of single systems to weak external forcings.  Nonetheless, the 
response of complex systems like arrays composed of many coupled units has also been
considered \cite{JBPM1992,MGC1995,gang,jung,lindner,schi,neiman,us06,us08,Pikovsky}.

The study of complex systems opens new possibilities with respect to those observed in single unit systems.
Besides the usual noise strength, the variation of other system parameters can
give rise to enhanced, nonmonotonic behaviors of some magnitudes characterizing the
system response to a weak externally applied time periodic force. Indeed,
nonmonotonic behaviors of pertinent quantifiers with the coupling strength parameter of the array
have been reported in the literature
\cite{JBPM1992,MGC1995,gang,jung,lindner,schi,neiman,us06,us08}.

The noise enhanced nonmonotonic response of driven systems with the number of units has been termed system size stochastic resonance (SSSR). It has been studied in several contexts. Several authors have considered the dependence of system properties on the number of elementary units $N$ in complex arrays of bistable elements \cite{Pikovsky,wio,lythe}. The dependence on the number of ion channels of the ion concentrations along cell membranes in biological Hodgkin-Huxley type models has been studied in \cite{schmidt,shuay}. In \cite{toral}, the authors analyze the optimization of neuron firing events in coupled excitable neuron units as the number of coupled elements is varied.  The phenomenon of SSSR has also been discussed within the context of modelling opinion formation in social collectivities \cite{tessone}. 

In this paper, we address the phenomenon of SSSR for arrays of globally coupled
noisy bistable units. We consider that the collective response of the whole
complex system is well represented by a single random variable. Its dynamics
is clearly constructed from the dynamics of the constituent elements of the array but  it shows
collective emergent properties. In Section \ref{model} we introduce a model in
terms of the individual units and the interactions among them, and we define the collective
variable and the quantifier that we will use to characterize the phenomenon. 
Next, in Section \ref{approx}, we discuss the possibility of a reduced
description of the collective variable dynamics in terms of an effective
Langevin equation, as put forward by Pikovsky et al. in
\cite{Pikovsky}. In  Section \ref{numerical}, we address the problem with
numerical simulation tools and discuss the results obtained. The paper ends with
some conclusions.
\section{The model}
\label{model}
The model considered in this work is the same as the one introduced by Desai
and Zwanzig in their influential paper \cite{deszwa}, except that we add a
time periodic force to the Langevin dynamics of each degree of freedom. This
model is also the one considered in \cite{Pikovsky}. The model consists of a
set of $N$ identical bistable units, each of them characterized by a variable
$x_i(t)\, (i=1,\ldots,N)$ satisfying stochastic evolution equations (in
dimensionless form) of the type 
\begin{equation}
\dot{x}_i=x_i-x_i^3+\frac{\theta}{N}\sum_{j=1}^N(x_j-x_i)+\sqrt{2D}\,\xi_i(t)+F(t),
\label{eq:lang}
\end{equation}
where $\theta$ is a coupling parameter and the term $\xi_i(t)$ represents a white noise with zero average and
$\left\langle \xi_i(t) \xi_j(s) \right\rangle = \delta_{ij}\delta
(t-s)$. The external driving force is periodic in $t$, $F(t)=F(t+T)$.
 An
alternative  formulation of the dynamics is in terms of the
Fokker-Planck equation for the joint probability density
$f_N(x_1,\ldots,x_N,t)$,
\begin{equation}
\label{lfpe} \frac{\partial f_N}{\partial t}=\sum_{i=1}^N
\frac{\partial}{\partial x_i}\Big ( \frac{\partial U}{\partial
x_i} f_N\Big  )+D\sum_{i=1}^N \frac{\partial^{2}f_N}{\partial
x_i^{2}},
\end{equation}
where $U$ is a time dependent potential energy relief,
\begin{eqnarray}
\label{dzmodel} 
U(x_1,\ldots,x_N,t)&=&\sum_{i=1}^N V(x_i) + \frac
{\theta}{4N} \sum_{i=1}^N \sum_{j=1}^N \left( x_i-x_j\right) ^{2}\nonumber \\
&& +F(t) \sum_{i=1}^N x_i
\end{eqnarray}
with
\begin{equation}
\label{bist} V(x)= \frac {x^4}{4}- \frac {x^2}2.
\end{equation}

We are interested in the properties of a collective variable, $S(t)$, defined as
\begin{equation}
S(t)=\frac 1N \sum_{j=1}^N x_j(t),
\end{equation}
characterizing the chain as a whole. Even though the set $x_i(t)$ is a
non-stationary $N$-dimensional Markovian process, $S(t)$ is not, in general, a
$1$-dimensional Markovian process.

Several magnitudes showing enhanced nonmonotonic behaviors as a system parameter is varied have been used as SR quantifiers \cite{GHJM1998}. 
In this work, we will use the
signal-to-noise ratio of the collective variable $R_\mathrm{out}$ as the
relevant quantifier.
In order to evaluate it, we need the knowledge of the one-time correlation
function of the collective variable defined as
\begin{equation}
C(\tau)=\frac 1T \int_0^T \,dt\; \langle S(t)S(t+\tau)\rangle_\infty ,
\end{equation}
The notation $\langle \ldots \rangle$ indicates an average over the
noise realizations and the subindex $\infty$ indicates the long time
limit of the noise average, i. e., its value after waiting for $t$
long enough that transients have died out. As shown in our previous
work \cite{us06}, it is convenient to split $C(\tau)$ into two parts, 
\begin{equation}
C(\tau)=C_\mathrm{coh}(\tau)+C_\mathrm{incoh}(\tau).
\end{equation}
The coherent part, $C_\mathrm{coh}(\tau)$, given by
\begin{equation}
C_\mathrm{coh}(\tau)=\frac 1T \int_0^T \,dt\; \langle S(t)\rangle_\infty \langle
S(t+\tau)\rangle_\infty ,
\end{equation}
is periodic in $\tau$
with the period of the driving force. The incoherent part,
$C_\mathrm{incoh}(\tau)$ arising from the fluctuations of the output $S(t)$
around its average value, decays to zero as $\tau$ increases. 

The quantifier $R_\mathrm{out}$ is defined as
\begin{equation}
\label{snr} R_\mathrm{out} =\lim_{\epsilon \rightarrow 0^+}\frac {
\int_{\Omega-\epsilon}^{\Omega+\epsilon} d\omega\;
\tilde{C}(\omega)}{\tilde{C}_\mathrm{incoh}(\Omega)}=\frac {
\tilde{C}_\mathrm{coh}(\Omega)}{\tilde{C}_\mathrm{incoh}(\Omega)} ,
\end{equation}
where $\Omega $ is the fundamental frequency of the driving force
$F(t)$, $\tilde{C}_\mathrm{coh}(\Omega)$ is the corresponding
Fourier coefficient in the Fourier series expansion of
$C_\mathrm{coh}(\tau)$, and $\tilde{C}_\mathrm{incoh}(\Omega)$ is
the Fourier transform at frequency $\Omega$ of
$C_\mathrm{incoh}(\tau)$.

\section{Approximate dynamics for $S(t)$}
\label{approx}
\begin{figure}
\includegraphics[width=8cm]{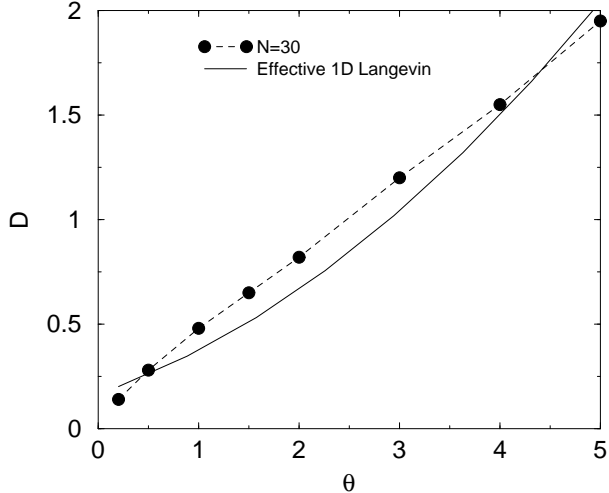}%
\caption{The solid line represents  $(2+\theta)^2-24D=0$. For points above it, the coefficient $b$ is a complex number and
  the effective Langevin equation (\ref{Eq10}) is not valid. The dotted line joining black dots
  marks a numerically obtained transition line such that, above this line, the equilibrium distribution of the
  global variable is always monomodal and below it $P^{\mathrm{eq}}(s)$ is
  multimodal (see the final paragraph of Section \ref{numerical}). The data depicted were obtained for a system with $N=30$ units, but those values are very independent of the system size as long as it remains finite.}
\label{FIG1}
\end{figure}

The dynamics of the collective variable follows from its definition and the
dynamics of the individual degrees of freedom. But, as $S(t)$ is not Markovian
in general, there is no reason why it should satisfy a 1-dimensional closed Langevin equation with
effective drift and diffusion coefficients. For the same reason, there is no guarantee that its associated
probability density $P(s,t)$ would satisfy a Fokker-Planck equation. On the
other hand, having an approximate description in terms of a 1-dimensional stochastic
process, would certainly be a helpful tool for the understanding of the
dynamical behaviors. It is then not surprising that several approximate
descriptions in a reduced 1-dimensional space for the collective variable
have been proposed in the literature \cite{Pikovsky,cubero}.  In
\cite{Pikovsky}, Pikovsky et al. used the Gaussian truncation of an infinite
hierarchy of equations for the cumulants and a slaving principle to construct
an effective $1$-dimensional Langevin equation. In the absence of external
driving, its explicit form is
\begin{equation}
 \dot S = aS-bS^3+\sqrt{\frac {2D}N}\chi(t)
\label{Eq10}
\end{equation}
with $\left\langle \chi(t) \chi(s) \right\rangle = \delta
(t-s)$ and the coefficients $a$ and $b$ given by
\begin{eqnarray}
 a&=&1+0.5(\theta-1)-0.5\sqrt{(\theta-1)^2 + 12 D}; \nonumber \\
 b&=&\frac {4a}{2-\theta + \sqrt{(2+\theta)^2-24D}}.
\label{Eq11}
\end{eqnarray}
Unfortunately, the coefficient $b$ does not exist for all ranges of the
$\theta$ and $D$ values. In Fig.\ (\ref{FIG1}) the continuous line represents  $(2+\theta)^2-24D=0$ containing the function 
appearing in the definition of $b$. This line separates
regions where the effective Langevin equation in Eq.\ (\ref{Eq10}) either
exists or not. Below that line, the $b$ coefficient is
real and the Langevin equation, Eq.\ (\ref{Eq10}), exists, while for points above the line $b$ is complex, and
therefore, the effective Langevin equation is not valid.

In general, $R_\mathrm{out}$ can not be evaluated analytically. But if we
accept that $S(t)$ satisfies a Langevin equation like Eq.\ (\ref{Eq10}) with
an extra term corresponding to the driving force added to its right hand side
and this extra term is very weak, then a linear response theory might provide
a valid approximation to analyze the system behavior. The problem is then
similar to that of SR in a single bistable potential and we can apply well
known arguments to obtain analytical
approximations \cite{Pikovsky,JungHanggi} for $R_\mathrm{out}$ within the limits of linear response theory. This is the strategy
followed by Pikovsky and coworkers in \cite{Pikovsky}. They find SSSR for weak
amplitude driving forces  and some ranges of noise strength and coupling
parameter values. Due to the difficulties with Eq.\
(\ref{Eq10}) discussed above, it is not clear whether the phenomenon exists for all parameter values or it is restricted to some
particular regions of the parameter space. It is also interesting to relate
the possible SSSR to the stochastic resonant effects in coupled arrays
reported in other works \cite{neiman,us06,us08} when the noise or the coupling
constant are varied.

\section{Numerical results}
\label{numerical}
The lack of a reliable effective Langevin approximation valid for all regions
of parameter space forces us to consider numerical simulations. Following the
procedure indicated in \cite{casgom03}, we have numerically solved the
Langevin equations for $x_i(t)$ in Eq.\ (\ref{eq:lang}) for very many noise
realizations. The numerical solution of the $N$-dimensional process is used to
collect information about $S(t)$ and construct the magnitudes that we need. In
particular we will estimate numerically the long-time limit of the first two
cumulant moments, the correlation function with its coherent and incoherent
parts and histograms to estimate the probability density. The 
estimation of $R_\mathrm{out}$ is then a matter of numerically performing the
quadratures indicated in its definition (see Eq.\ (\ref{snr})).

Let us consider a rectangular driving force: $F(t)=A$ ($F(t)=-A$) if $t\in [n
T/2,(n+1)T/2)$ with $n$ even (odd). In what follows, we will always take the
amplitude $A=0.05$ and the fundamental frequency $\Omega=0.05$. When such a
force is applied to a single isolated noisy bistable, the response of the
system is very well described by a linear response function, as the dynamical
effects arising from the distortion of the bistable potential induced by the
driving are within the limits of a small perturbation theory.

\subsection{Strong coupling}
\begin{figure}
\includegraphics[width=7cm]{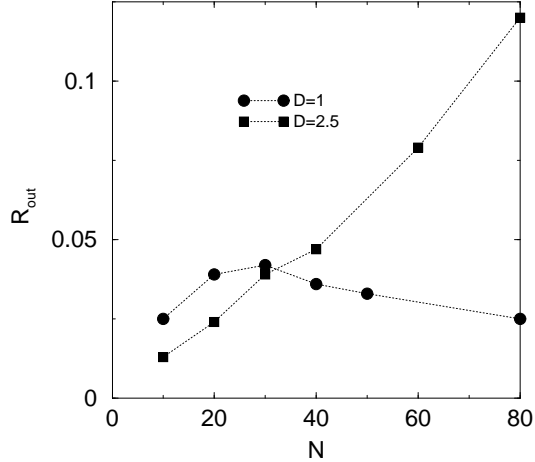}
\caption{The signal-to-noise ratio of the collective output $R_{\mathrm{out}}$ as a
  function of $N$ for two different values of the noise intensity $D$. In both
  cases $\theta=5.5$. The driving amplitude is $A=0.05$ and the fundamental
frequency $\Omega=0.05$.}
\label{FIG2}
\end{figure}

\begin{figure}
\includegraphics[width=8cm]{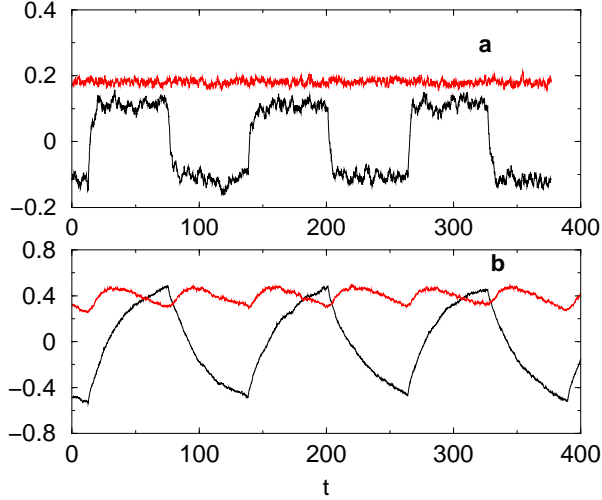}%
\caption{(Color online) Time evolution of the first two cumulant moments of $S(t)$ in a system with $N=30$ strongly coupled ( $\theta=5.5$) units driven by a rectangular force with $A=0.05$ and $\Omega=0.05$, for $D=2.5$ (a) and $D=1$ (b).}
\label{FIG3} 
\end{figure}

\begin{figure}
\includegraphics[width=8cm]{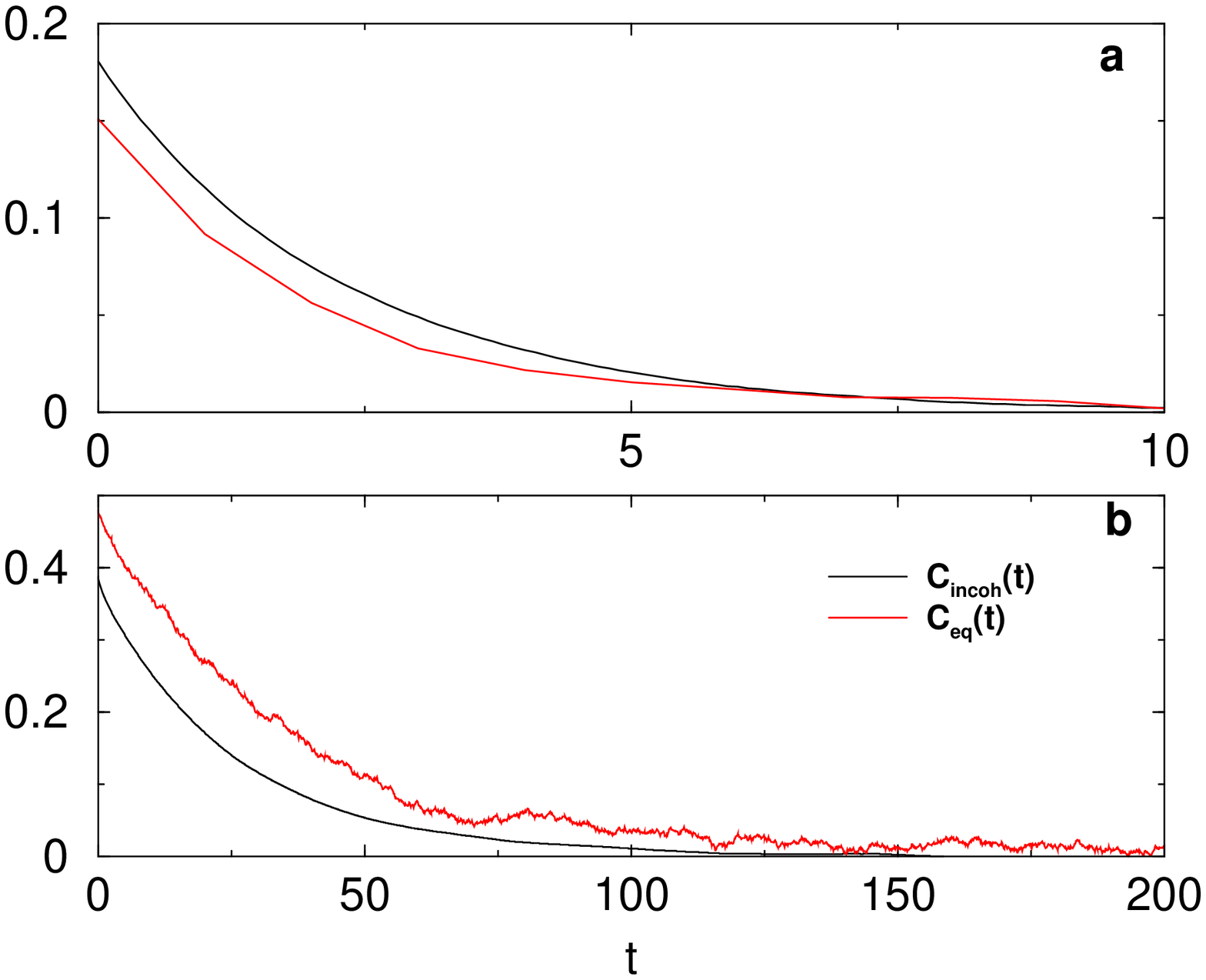}%
\caption{(Color online) Time evolution of the incoherent part of the correlation 
function (black line) and the equilibrium time correlation function (red line) of the collective variable in a system with $N=30$ strongly coupled ( $\theta=5.5$) units for $D=2.5$ (a), and $D=1$ (b). The driving force is rectangular with $A=0.05$ and $\Omega=0.05$.}
\label{FIG4}
\end{figure}

First we study the case of parameter values within the range considered in
\cite{Pikovsky}. In Fig.\ (\ref{FIG2}), the black circles correspond to the
numerically obtained values of $R_\mathrm{out}$ as the number of particles $N$
of the array is varied and $\theta=5.5$, $D=1$. The existence of SSSR is evident, in consonance
with the results reported in \cite{Pikovsky}. It should also be pointed out
that the values of $R_\mathrm{out}$ are rather small, even for arrays with a
size where $R_\mathrm{out}$ reaches its peak value. To understand why, we have
analyzed the time behavior of the first two cumulants of the collective
variable as well as the incoherent part of the fluctuations about its average.
In Fig.\ (\ref{FIG3}b) we depict the time evolution of the
first two cumulant moments for the parameter values $\theta=5.5$, $D=1$. Even
though the amplitude of the average output is about five times bigger than
that of the driving force, the output signal is very noisy with a second
cumulant much larger than the noise strength $(2D)/30\sim 0.066$ associated to
the effective Langevin equation, Eq.\ (\ref{Eq10}). The incoherent
fluctuations of the collective variable are depicted in 
Fig.\ (\ref{FIG4}b). They are large and long-lasting. Consequently, the
numerator in the fraction defining $R_\mathrm{out}$ (see Eq.\ (\ref{snr}))
should not be very large, while its denominator has a substantial value. The
smallness of the $R_\mathrm{out}$ values is not surprising. 

The results of
Pikovsky et al. \cite{Pikovsky} rely on the validity of the linear response
approximation. Within this limit, one assumes that the incoherent part of the
output correlation function in a driven system can be safely approximated by
the equilibrium correlation function of the same system in the absence of
driving force. It is clear from Fig.\ (\ref{FIG4}b) that such an
approximation is not too bad. Then, for the parameter values $\theta=5.5$,
$D=1$, the linear response function can be used and as this function has a
nonmonotonic behavior with $N$, one finds SSSR.

Let us now analyze what happens if we increase the noise strength to $D=2.5$,
while keeping all the other parameter values fixed. The black squares in Fig.\
(\ref{FIG2}) indicate that $R_\mathrm{out}$ increases monotonically with the
size of the system. Thus, there is no SSSR for
this noise strength even though the coupling parameter is still strong. 
Note that for these noise strength and coupling parameter
values, the effective Langevin equation in Eq.\ (\ref{Eq10}) does not exist.
On the other hand, the numerical results
depicted in Figs.\ (\ref{FIG3}a) and (\ref{FIG4}a) show that the second
cumulant is reduced with respect to its value for $D=1$, and that the incoherent
part of the fluctuations in the driven system matches well the
equilibrium fluctuations. Taking also into account the smallness of the
driving amplitude, a linear response approximation should still provide a reliable
description of the response of the system for large noises and strong couplings,
even though SSSR does not exist in that region. 

\begin{figure}
\includegraphics[width=8cm]{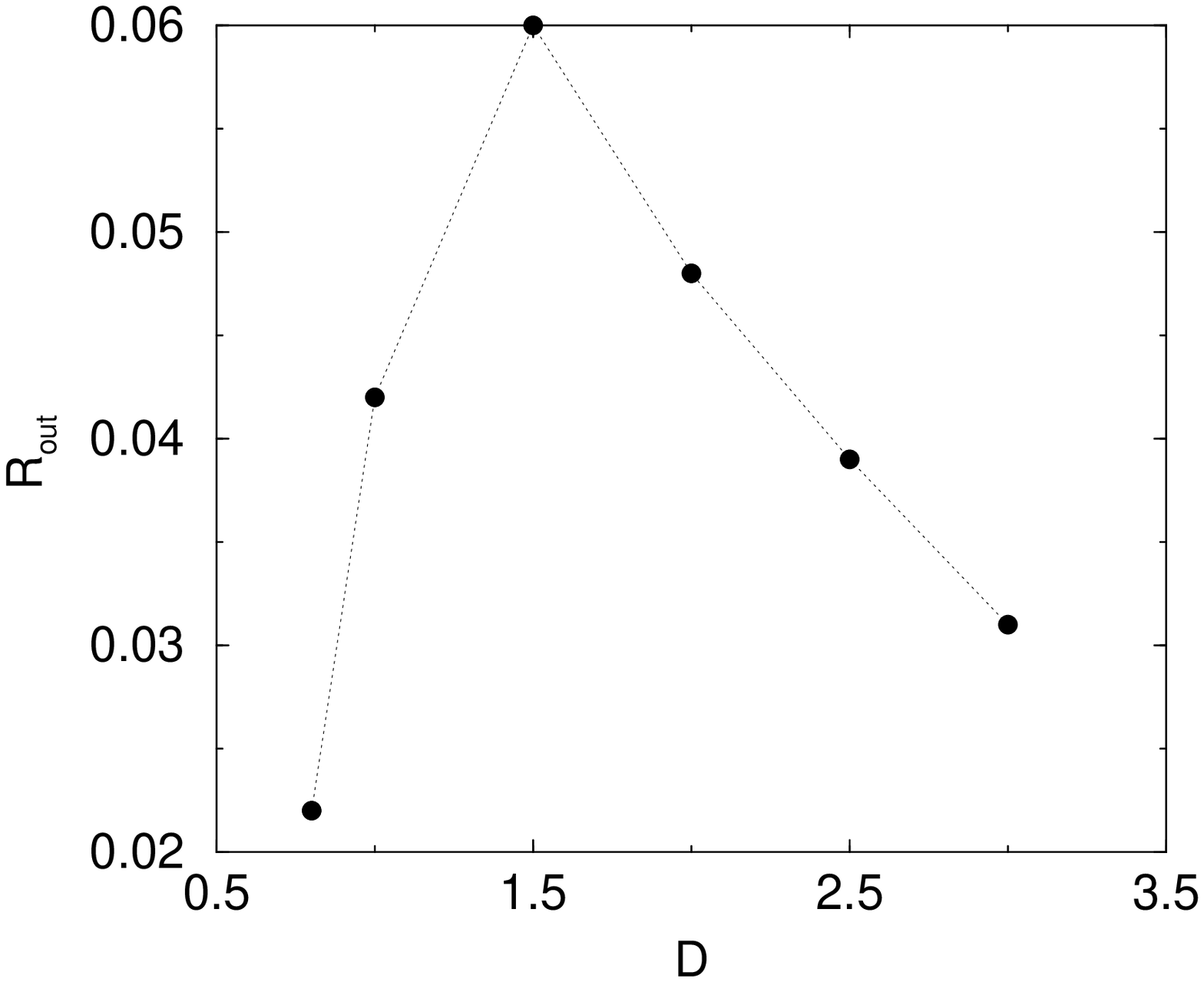}%
\caption{Non-monotonic behavior of $R_{\mathrm{out}}$ with the
  noise strength $D$ for a system with $N=30$ units with global coupling strength
  $\theta=5.5$. The driving force is rectangular with $A=0.05$ and $\Omega=0.05$. }
\label{FIG5}
\end{figure}

It is worth noting that a nonmonotonic behavior of $R_\mathrm{out}$
with the noise strength $D$, typical of stochastic resonance, does exist for strong coupling. 
An example is depicted in Fig.\ (\ref{FIG5}) for a system of $N=30$ elements with coupling strength $\theta=5.5$,
driven by a weak rectangular force with $A=0.05$ and $\Omega=0.05$. As expected, the values of $R_\mathrm{out}$ are quite small.

\subsection{Weak coupling}

\begin{figure}
\includegraphics[width=8cm]{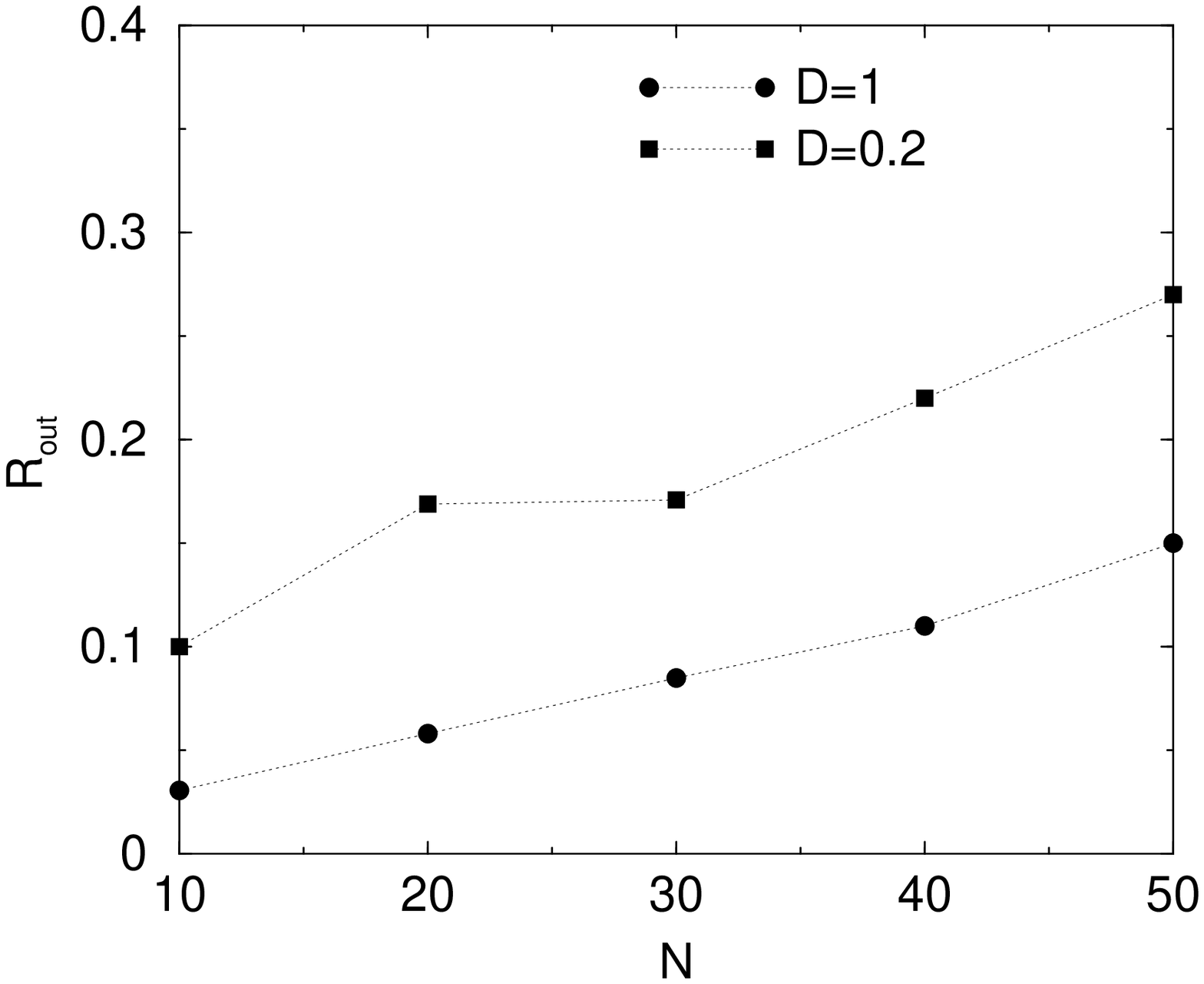}%
\caption{Monotonic behavior of $R_{\mathrm{out}}$ with the number
  of units $N$. The data corresponds to $\theta=0.5$ and two different
  values of $D$. The driving force is rectangular with $A=0.05$ and $\Omega=0.05$.}
\label{FIG6}
\end{figure}

Let us now consider cases with a much weaker coupling parameter. We will take
$\theta=0.5$. We will consider two values of the noise strength: $D=1$ and
$D=0.2$. In Fig.\ (\ref{FIG6}) we depict the behavior of $R_\mathrm{out}$
with the system size. Clearly, there is no SSSR as the value of $R_\mathrm{out}$ increases with
$N$ for both noise values. 

\begin{figure}
\includegraphics[width=8cm]{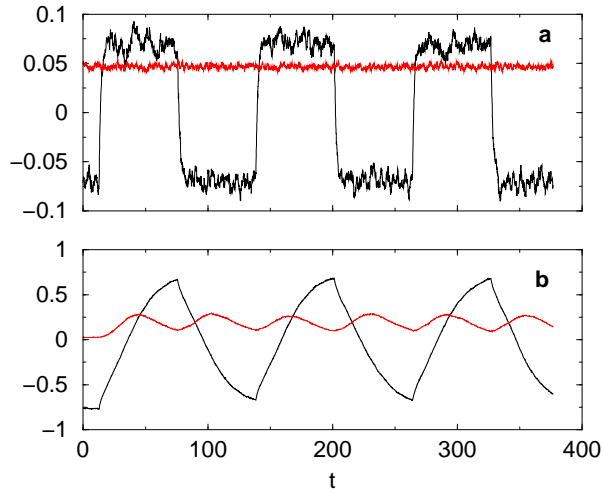}%
\caption{(Color online) Time evolution of the first two cumulant moments of $S(t)$ in a system with $N=30$ weakly coupled ($\theta=0.5$) units driven by a rectangular force with $A=0.05$ and $\Omega=0.05$, for $D=1$ (a) and $D=0.2$ (b)}
\label{FIG7}
\end{figure}

The behavior of the first two cumulant moments of the collective variable for
a driven system with $N=30$ units and coupling parameter value $\theta=0.5$ is
depicted in Fig.\ (\ref{FIG7}) for two values of the noise strength: $D=0.2$
in the lower panel, and $D=1$ in the upper one. It is to be noted the large increase of 
the output amplitude relative to the weak driving amplitude for the $D=0.2$ case, while in the higher 
noise situation, the amplitude amplification is small. The second cumulant plots indicate that
the output noise level is also very much increased in the low $D$ case, while it is kept very small in
for the higher noise.    

\begin{figure}
\includegraphics[width=7cm]{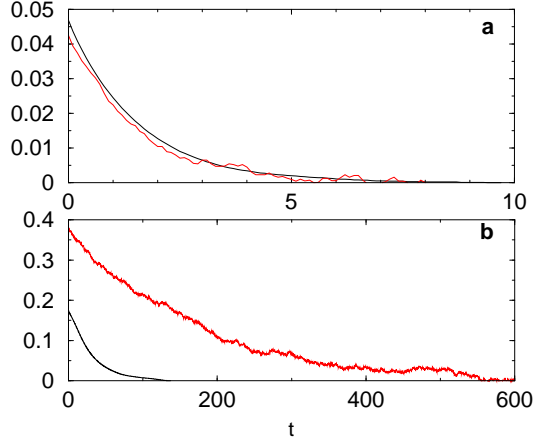}%
\caption{(Color online) Time evolution of the incoherent part of the correlation 
function (black line) and the equilibrium time correlation function (red line) of the collective variable in a system with $N=30$ weakly coupled ($\theta=0.5$) units for $D=1$ (a), and  $D=0.2$ (b). The driving force is rectangular with $A=0.05$ and $\Omega=0.05$.}
\label{FIG8}
\end{figure}

The time dependence of the
incoherent part of the correlation function of the collective variable for the
driven system and the equilibrium correlation function of the same variable
are shown in Fig.\ (\ref{FIG8}), for $N=30$, $\theta=0.5$ $A=0.05$,
$\Omega=0.05$ and two noise values. For the higher noise value, $D=1$, the incoherent fluctuations in the driven
system are practically identical to the equilibrium correlation function. This feature indicates that
a linear response theory should then be an excellent approximation to
analyze the response of the system to weak driving forces, when the noise strength is large even in a 
weak coupling situation. On the other hand, for the low noise strength case $D=0.2$,
the situation is different. The time behavior of $C_\mathrm{incoh}(t)$ is very
different from the time behavior of $C_\mathrm{eq}(t)$.  The correlations
decay much faster in the driven system than in the absence of driving and from
a smaller initial value. This indicates that the driving field, albeit weak,
introduces a very strong distortion of the fluctuation dynamics of
the collective variable during each cycle of the external driving, thus
rendering inadequate the ideas behind a linear response approximation. This
fact has been indeed analyzed previously by us in connection with SR in arrays of interacting particles
\cite{CCGCM2007,MGCCC2009}. We have referred to this regime as a non-linear SR
regime, characterized by the control of the fluctuations by the external
driving.

\begin{figure}
\includegraphics[width=7cm]{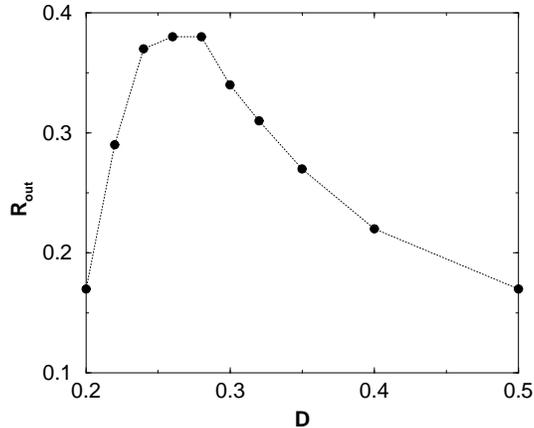}
\caption{ Non-monotonic behavior of $R_{\mathrm{out}}$ with $D$ 
for $N=30$ weakly coupled ($\theta=0.5$) units. The driving force is rectangular with $A=0.05$ and $\Omega=0.05$.}
\label{FIG9}
\end{figure}

In Fig.\ (\ref{FIG9}) we depict the nonmonotonic behavior of $R_\mathrm{out}$
with $D$ for an array of weakly coupled ($\theta=0.5$) $N=30$ units, driven by
a rectangular force with $A=0.05$ and $\Omega=0.05$.  The values of
$R_\mathrm{out}$ are substantially larger than the ones obtained for the
stronger coupling constant (compare with Fig.\ \ref{FIG5}).

To help rationalize the results above, it is useful to consider the
equilibrium distribution function of the collective variable
$P^\mathrm{eq}(s)=\langle \delta(s-S(t)) \rangle$. This function can be
estimated by constructing histograms from the long time results of the
numerical simulations of Eq.\ (\ref{eq:lang}) with $F(t)=0$. The data reveals
the existence of a transition line separating different regions in the
$D$-$\theta$ plane.  The line joining the black circles in Fig.\ (\ref{FIG1})
corresponds to the transition line for a system with $N=30$ particles. The
location of the line for systems with other number of particles do not differ
substantially from the one plotted here. For parameter values corresponding to
points above the line, the equilibrium distribution is always monomodal and
centered around $s=0$, while for points below the line, it is bimodal with one
minimum at $s=0$ and two maxima at values $\pm s_0$. The location of the
minima depends on the values of the system parameters. Thus for a fixed $N$
and $\theta$, as the noise value is increased, the equilibrium distribution of
the system would change shape from bimodal to monomodal.  The nonmonotonic behavior of
$R_\mathrm{out}$ with $D$ (SR) observed in both Figs.\ (\ref{FIG5}) (for strong
coupling) and (\ref{FIG9}) (for weak
coupling) is associated the change in the shape of the equilibrium
distribution function. By contrast, SSSR is 
restricted to a reduced region of parameter space involving rather high values of
the coupling strength and relatively low values of the noise strength without involving 
a change in shape of $P^\mathrm{eq}(s)$. SSSR is then connected to conditions 
such that a bimodal equilibrium distribution is only slightly distorted by the driving field. 
Note that even though a linear response approach might yield a reliable description of the dynamics, 
as it happens for points above the transition line, the existence of SSSR is not guaranteed.
In this sense, SSSR 
does not seem to be such a general effect as the usual SR nonmonotonic behavior with
the noise strength.

\section{Conclusions}
In this work, we have analyzed the response of globally coupled arrays of noisy bistable systems driven by weak time
periodic forces. We focus on the dynamics of a 1-dimensional collective variable characterizing the array as a whole.
The lack of an adequate Langevin equation describing the reduced dynamics of the collective variable, prompts us to use
numerical simulations involving all the individual dynamics. 

Our results indicate that SSSR exists only in limited regions of parameter space:
for relatively low noise values and strongly coupled arrays. By looking at the behavior of the fluctuations in those regions, we find that the
system response is adequately well described by a linear response approximation. Furthermore, the reduced dynamics 
can be well described by an effective Langevin equation. 

Even for strongly coupled arrays, SSSR does not exist when the noise values are large. The effective Langevin equation
proposed in \cite{Pikovsky} ceases to be valid. On the other hand, the equilibrium probability of the collective variable is,
in this region of parameter values, monomodal. Thus, the effect of a weak applied external force on
the collective dynamics should still well described by a linear response function which has a monotonic behavior with $N$.
  
For weak coupling, the SSSR disappears. Regardless of the noise strength, $R_\mathrm{out}$ monotonically increases with the number of particles in the array. For large noise values, $P^\mathrm{eq}(s)$ is monomodal. Thus, a linear response approximation should be a reliable description of the system response. On the other hand, for low noise values, the influence of the applied force on the fluctuations dynamics is so intense that the system operates in a non-linear regime. 

Our results demonstrate that in the strong coupling regime, there is a typical SR behavior with the noise strength $D$, when the system size $N$ and the coupling constant $\theta$ are kept fixed. The $R_\mathrm{out}$  values are, nonetheless, very small. This is in sharp contrast with the large enhancement of the SR effects observed in the weak coupling regime. 

The construction of effective reduced 1-dimensional dynamics for the relevant variable is very useful. The approaches followed in \cite{Pikovsky,cubero} are unfortunately limited to restricted regions of parameter space. We are currently working on obtaining approximate Langevin equations for the collective variable in driven systems valid for all regions in parameter space, along the lines initiated in \cite{honisch}.

\begin{acknowledgments}
We acknowledge the support of the Ministerio de Ciencia e Innovaci\'on
of Spain (FIS2008-04120)
\end{acknowledgments}


\begin{thebibliography}{100}
\bibitem{GHJM1998} L. Gammaitoni, P. H\"{a}nggi, P. Jung and
  F. Marchesoni. Rev. Mod. Phys. {\bf 70}, 223 (1998).
\bibitem{BM2005} R. L. Badzey and P. Mohanty, Nature, {\bf 437}, 995 (2005).
\bibitem{JBPM1992} P. Jung, U. Behn, E. Pantazelou and F. Moss, Phys. Rev. A
  {\bf 46}, R1709 (1992).
\bibitem{MGC1995} M. Morillo, J. G\'omez-Ord\'o\~nez and J. M. Casado,
  Phys. Rev. E {\bf 52}, 316 (1995).
\bibitem{jung} P. Jung and G. Mayer-Kress, Phys. Rev. Lett. \textbf{74}, 2130 (1995);
\bibitem{lindner} John. F. Lindner, Brian K. Meadows, William L. Ditto, Mario E. Inchiosa, and Adi R. Bulsara,   Phys. Rev. Lett.  \textbf{75}, 3 (1995).
\bibitem{schi} Lutz Schimansky-Geier and Udo Siewert, in \textit{Stochastic Dynamics} (Springer, Berlin 1997) p. 245.
\bibitem{gang} H. Gang, H. Haken, and X. Fagen, Phys. Rev. Lett. \textbf{77}, 1925 (1996).
\bibitem{neiman} A. Neiman, L. Schimansky-Geier and F. Moss, Phys. Rev. E \textbf{56}, R9 (1997).
\bibitem{us06} J. M. Casado, J. G\'omez-Ord\'o\~nez and M. Morillo, 
Phys. Rev. E \textbf{73}, 011109 (2006).
\bibitem{us08} Manuel Morillo, Jos\'e G\'omez Ord\'o\~nez and Jos\'e M. Casado, Phys. Rev. E \textbf{78}, 021109 (2008).
\bibitem{Pikovsky} A. Pikovsky, A. Zaikin and M. A. de la Casa, Phys.
Rev. Lett. \textbf{88}, 050601 (2002).
\bibitem{wio} B. von Haeften, G. Iz\'us, and H. S. Wio, Phys. Rev. E, \textbf{72}, 021101 (2005).
\bibitem{lythe} Mario Castro and Grant Lythe, SIAM J. Applied Dynamical Systems, \textbf{7}, 207 (2008).
\bibitem{schmidt} G. Schmidt, I. Goychuk, and P. H\"anggi, Europhys. Lett. \textbf{56}, 22 (2001); Phys. Biol. \textbf{1}, 61 (2004).
\bibitem{shuay} P. Jung and J. W. Shuay, Europhys. Lett. \textbf{56}, 29 (2001); Phys. Rev. Lett. \textbf{88}, 068102 (2003).
\bibitem{toral} R. Toral, C. Mirasso, and J. Gunton, Europhys. Lett. \textbf{61}, 162 (2003).
\bibitem{tessone} Claudio J. Tessone, Ra\'ul Toral, Physica A \textbf{351}, 106 (2005).
\bibitem{deszwa} Rashmi C. Desai and  Robert Zwanzig, J. Stat. Phys.
\textbf{19}, 1 (1978).
\bibitem{cubero} David Cubero, Phys. Rev. E \textbf{77}, 021112 (2008).
\bibitem{JungHanggi} P. Jung and P. H\"anggi, Phys. Rev. A \textbf{44}, 8032
(1991).
\bibitem{casgom03} Jes\' us Casado-Pascual, Claus Denk, Jos\' e G\' omez-
Ord\'o\~nez, Manuel Morillo, and Peter H\"anggi, Phys. Rev. E \textbf{68},
061104 (2003).
\bibitem{CCGCM2007} D. Cubero, J. Casado-Pascual, J. G\'omez-Ord\'o\~nez,
  J. M. Casado and M. Morillo, Phys. Rev. E {\bf 75}, 062102 (2007).
\bibitem{MGCCC2009}M. Morillo, J. G\'omez-Ord\'o\~nez, J. M. Casado,
  J. Casado-Pascual and D. Cubero, Eur. Phys. J. \textbf{69}, 59 (2009). 
\bibitem{honisch} Jos\' e G\' omez-Ord\'o\~nez, Jos\'e M. Casado, Manuel Morillo, Christoph Honisch and Rudolf Friedrich, arXiv/cond-matt. 0905.1564, (2009).

\end{thebibliography}
\end{document}